\documentclass[runningheads]{llncs}
\usepackage{amsmath,amssymb,amsfonts}
\usepackage[T1]{fontenc}
\usepackage{algorithmic}
\usepackage{graphicx}
\usepackage{textcomp}
\usepackage{xcolor}
\usepackage{tikz}
\usepackage{url}
\usepackage[linesnumbered,ruled,vlined]{algorithm2e}
\usepackage{pgfplots}
\usepackage{hyperref}

\usepackage{booktabs}
\usepackage{multirow}
\usepackage{numprint}
\npstyleenglish

\newtheorem{Theorem}{Theorem}
\newtheorem{Definition}{Definition}
\newtheorem{Remark}{Remark}
\newtheorem{Lemma}{Lemma}

\begin{document}
\title{Cryptanalysis of a Lattice-Based PIR Scheme for Arbitrary Database Sizes}

\author{Svenja Lage}

\institute{Institute of Communications and Navigation\\
German Aerospace Center (DLR)\\
Oberpfaffenhofen, Germany\\
svenja.lage@dlr.de}
\maketitle              
\begin{abstract}

Private Information Retrieval (PIR) schemes enable users to securely retrieve files from a server without disclosing the content of their queries, thereby preserving their privacy. In 2008, Melchor and Gaborit proposed a PIR scheme that achieves a balance between communication overhead and server-side computational cost. However, for particularly small databases, Liu and Bi identified a vulnerability in the scheme using lattice-based methods. Nevertheless, the rapid increase in computational cost associated with the attack limited its practical applicability, leaving the scheme's overall security largely intact. In this paper, we present a novel two-stage attack that extends the work of Liu and Bi to databases of arbitrary sizes. To this end, we employ a binary-search-like preprocessing technique, which enables a significant reduction in the number of lattice problems that need to be considered. Specifically, we demonstrate how to compromise the scheme in a matter of minutes using an ordinary laptop. Our findings are substantiated through both rigorous analytical proofs and comprehensive numerical experiments.

\end{abstract}

\section{Introduction}
Private Information Retrieval (PIR) schemes enable users to access data from a server while preventing the server from inferring the specific file being requested. Thus, PIR schemes maintain the confidentiality of users' queries. However, this process introduces significant computational overhead. In the worst-case scenario, the user would have to download the entire database to securely conceal the query. Although the request remains hidden, this solution is impractical for large databases with many files. However, as demonstrated in \cite{PIR}, this is the only method to achieve information-theoretical secrecy.\\

As a more practical alternative, the concept of computationally secure PIR schemes has emerged, providing a trade-off between privacy and communication cost. The majority of PIR schemes focus on optimizing communication cost while ensuring a predetermined level of security. In 2008, Melchor and Gaborit \cite{OriginalScheme} introduced a PIR scheme that addressed both communication costs and the often-overlooked computational burden on the server side. The authors aimed at balancing both aspects to develop a secure and efficient PIR scheme. However, in 2016, Liu and Bi \cite{Attack} discovered a linear relationship between the queries and the secret noise matrix holding the information about the desired file index, which can be exploited using a lattice reduction approach. However, their attack requires performing an amount of lattice reductions which grows linearly with the number of files in the database, rendering it infeasible for large databases.\\

In this paper, we propose a two-stage attack that compromises the scheme for databases of all sizes. Our approach involves adding a preprocessing step that significantly reduces the number of lattice reductions required, allowing the original attack to be applied subsequently. The time complexity of our improved attack grows only logarithmically with the number of files in the database, making it possible to consider arbitrarily large databases. 
After introducing the original scheme and the initial attack, Section \ref{ImprovedAttack} describes our enhanced attack and proves its correctness. Subsequently, a numerical analysis is presented to underscore the efficiency of the attack.

\section{Preliminaries}\label{Preliminaries}
We first introduce some definitions and results needed for the subsequent sections. For a prime $p$, let $\mathbb{Z}_p$ be the finite field with $p$ elements and denote the set of all $n\times m$ matrices over $\mathbb{Z}_p$ by $M_{n,m}(\mathbb{Z}_p)$. For the special case of square matrices, we write $M_{n}(\mathbb{Z}_p)$. The subset of all non-singular $n\times n$ matrices over $\mathbb{Z}_p$ is denoted by $GL_{n}(\mathbb{Z}_p)$. For any $A\in M_{n,m}(\mathbb{Z}_p)$, we use $A|_{[u:v,:]}$ to refer to the submatrix of $A$ consisting of rows $u$ through $v$ and $A|_{[:,u:v]}$ to refer to the submatrix of $A$ consisting of columns $u$ through $v$.\\ 

A lattice $L\subset\mathbb{R}^d$ is formally defined as a discrete subgroup of $\mathbb{R}^d$. For a set of linearly independent vectors $B=[b_1,\ldots,b_n]$ with $b_i\in\mathbb{R}^d$, the lattice generated by $B$ is defined as
\begin{align}\label{lattice}
    L(B)= \{y\in \mathbb{R}^d : y=Bx & \textrm{ for some }x\in\mathbb{Z}^n\}.
\end{align}
The matrix $B$ is called a lattice basis, while $n$ is referred to as the rank of the lattice.\\
In this work, we focus on $q$-ary lattices, which are defined as
\begin{align*}
     L_q(B)= \{y\in \mathbb{Z}^d : y=Bx \hspace{-0.2cm}\mod q & \textrm{ for some }x\in\mathbb{Z}_q^n\}
\end{align*}
for a fixed $q\in\mathbb{N}$ and $B\in\mathbb{Z}_q^{d\times n}$. The lattice basis in $q$-ary lattices is defined as the columns of the matrix $E = (B \,| \, qI_d)$, where $I_d$ denotes the $d$-dimensional identity matrix. This augmented matrix construction enables the lattice to simultaneously capture the linear combinations generated by $B$ and the modular reductions imposed by the parameter $q$. Utilizing the basis $E$, we can display the q-ary lattice as an ordinary lattice $L(E)$ as defined in (\ref{lattice}). Conversely, every lattice $L$ with $q\mathbb{Z}^d\subseteq L \subseteq\mathbb{Z}^d$ is an q-ary lattice.\\

In lattice theory, the Shortest Vector Problem (SVP) is fundamental and provides a basis for numerous cryptographic applications. It can be stated as follows.
\begin{Definition}[Shortest Vector Problem (SVP)]
    Given a basis $B$ of a lattice $L$ and a norm $||\cdot||$, find a non-zero vector $v\in L$ such that 
    \begin{align*}
        ||v|| =\min_{u\in L\setminus \{0\}} ||u||.
    \end{align*}
\end{Definition}
The SVP seeks the shortest non-zero lattice vector with respect to a given norm. Unless otherwise specified, in the following sections, any reference to a norm, denoted as $||\cdot||$, shall be interpreted as the Euclidean norm. The computational complexity of SVP depends on the problem's dimensionality and the quality of the lattice bases provided. However, in the Euclidean norm, the SVP is at least known to be NP-hard under randomized reductions \cite{Ajtai}. 

Closely related to the SVP is the Closest Vector Problem (CVP), which asks for the lattice vector nearest to a given target vector $t$. The CVP is known to be NP-hard in the Euclidean norm \cite{NPcomplete}. 

\begin{Definition}[Closest Vector Problem (CVP)]
    Given a basis $B$ of a lattice $L$, a target vector $t\notin L$ and a norm $||\cdot||$, find a vector $v\in L$ such that 
    \begin{align*}
        ||v-t|| =\min_{u\in L} ||u-t||.
    \end{align*}
\end{Definition}

In our subsequent attack, we are initially confronted with an instance of the Closest Vector Problem on a lattice $L$ and target vector $t$. Given the existence of more efficient solvers for Shortest Vector Problems, we leverage Kannan's embedding technique \cite{Kannan} to transform the CVP into a higher-dimensional SVP.  

Kannan's approach starts with a short basis $B \in \mathbb{Z}^{d \times d}$ of the lattice $L$, where a short basis is characterized by its reduced norm and nearly orthogonal vectors. If the initial basis $B$ does not already possess these desirable properties, a lattice reduction algorithm is utilized to transform $B$ into a short basis, thereby optimizing its form for subsequent computations. As a second step, an embedded lattice with basis
\begin{align*}
\begin{pmatrix}
B & t \\
0 & M \\
\end{pmatrix} \in \mathbb{Z}^{(d+1) \times (d+1)}
\end{align*}
is constructed for an embedding factor $M \in \mathbb{N}$. Calculating the shortest vector $(e,M)^T$ in the embedded lattice, we obtain the closest vector to $t$ by considering the vector $(t-e)$. According to \cite{StudyKannan}, the selection of $M$ has a significant impact on the difference between the shortest and the second shortest vector in the embedded lattice. When $M$ is excessively large, the gap between these vectors diminishes, thereby complicating the task of identifying the shortest vector. Conversely, the probability of Kannan's embedding technique failing to produce the closest vector to $t$ rises as $M$ decreases. As suggested by \cite{StudyKannan}, a balanced trade-off for $M$ is to select a value proportional to the expected value of $||e||$. However, each application necessitates a separate study of the optimal choice of $M$. \\

To estimate the number of lattice points within a certain distance to a given point, we employ the following definitions. The $d$-dimensional sphere of radius $R > 0$ centered at $x \in \mathbb{R}^d$ is denoted by $B_R(x)$ with
\begin{align*}
    B_R(x)= \{y\in\mathbb{R}^d : ||y-x||\leq R\}.
\end{align*}
Let $N_R(x)$ represent the number of integer points within $B_R(x)$, which is given by
\begin{align*}
    N_R(x)= |\{y\in\mathbb{Z}^d : ||y-x||\leq R\}|.
\end{align*}
Estimating $N_R(x)$ precisely is non-trivial. However, for $d \geq 4$, it can be shown \cite{Chamizo1995}, that $N_R(x)$ can be approximated by the volume of the $d$-dimensional sphere as
\begin{align*}
    N_R(x) = \frac{\pi^{\frac{d}{2}}}{\Gamma\left(\frac{d}{2}+1\right)}R^d +O(R^{d-2}),  
\end{align*}
where $\Gamma(\cdot)$ denotes the Gamma function and $O(\cdot)$ represents the big O notation. It is important to note that shifting the center $x$ between integers does not change $N_R(x)$. However, shifting $x$ to a non-integer point does indeed influence the number of integer points within the sphere, as demonstrated in \cite{Odlyzko1990}. \\

\section{The original PIR scheme}
The authors of \cite{OriginalScheme} introduced a lattice-based PIR scheme, which operates on a database consisting of $n$ files, each represented as an $L \times N$ matrix, $A_1, A_2, \ldots, A_n$ over $\mathbb{Z}_p$.  The user aims to retrieve the file indexed by $i_0 \in \{1, 2, \ldots, n\}$, while ensuring that no information about $i_0$ is disclosed to the server. In this scheme, the parameter $N$ is a crucial scheme parameter influencing the lattice size. Meanwhile, the parameter $L$ is chosen to be sufficiently large to allow for the encoding of all files within the database. 

\subsection{Query generation}
For each file in the database, the user generates a query $B_i$, $i=1,\ldots,n$, where $B_i \in M_{N,2N}(\mathbb{Z}_p)$ is calculated according to the following procedure. 

\begin{itemize}
    \item [1.] Define the parameters $l_0$ and $q$ as $l_0 = \lceil \log(nN) \rceil + 2$ and $q = 2^{2l_0-1}$, respectively. Additionally, a prime number $p$ is chosen such that $p > 2^{3l_0}$, thereby guaranteeing that $q$ remains small compared to $p$.
    \item [2.] Two random matrices, $M_1 \in GL_N(\mathbb{Z}_p)$ and $M_2 \in M_N(\mathbb{Z}_p)$, are generated, along with a random scrambling matrix $\Delta \in GL_{2N}(\mathbb{Z}_p)$. These matrices will be utilized in the construction of all query matrices $B_1, B_2, \ldots, B_n$ and must be stored for later use during information extraction.
    \item [3.]  For each $i=1,\ldots,n$ generate a random matrix $P_i\in  GL_{N}(\mathbb{Z}_p)$. In addition, randomly generate noise matrices $\epsilon_i\in\{-1,1\}^{N\times N}$. 
    \item [4.] For the index of interest $i_0$, modify the noise matrix by multiplying its diagonal entries with $q$. This step embeds the index of interest into the amplitude of the noise, which conceals the user's query. 
    \item [5.] For  each $i=1,\ldots,n$ compute the query matrix $B_i$ as 
    \begin{align*}
        B_i = [P_iM_1 \,| \, P_iM_2+\epsilon_i]\Delta .
    \end{align*}
\end{itemize}
Finally, the user transmits the request tuple $(B_1, \ldots, B_n, p)$ to the server, thereby initiating the privacy-preserving retrieval process.

\subsection{Answer encoding}
In the database, the files $A_1, \ldots, A_n$ are represented as $L \times N$ matrices, where each entry is an $l_0$-bit scalar. In response to a query, the server computes the $L \times 2N$ matrix $R$ as $R = AB$, where $A$ is the $L \times nN$ matrix obtained by horizontally concatenating the individual file matrices $A_1, \ldots, A_n$. Formally, $A$ is given by 
\begin{align*}
    A=[A_1|\ldots|A_n]\in M_{L,nN}(\mathbb{Z}_p).
\end{align*}
The matrix $B$ is the $nN \times 2N$ matrix constructed by vertically stacking the transposes of the query matrices $B_1, \ldots, B_n$, i.e.,
\begin{align*}
    B=[B_1^T|\ldots|B_n^T]^T\in M_{nN,2N}(\mathbb{Z}_p).
\end{align*}
The server performs the matrix multiplication over $\mathbb{Z}_p$ and returns $R$ as the response to the user's query.

\subsection{Information extraction}
Once the server encodes the query response, the user can extract the file $A_{i_0}$ from the matrix $R$.
To this end, the user adopts a row-wise approach. Specifically, for each row $V_i$, $i = 1, \ldots, L$, of $R$ the user employs the following procedure.

\begin{itemize}
    \item [1.] Since $\Delta$ is invertible, calculate $\Delta^{-1}$ and unscramble $V_i$ as
    \begin{align*}
        V_i' = V_i\Delta^{-1}. 
    \end{align*}
    \item [2.] The user partitions the modified row $V_i'$ into two segments: the undisturbed part $V_i'(U)$, comprising the first $N$ columns, and the disturbed part $V_i'(D)$, consisting of the remaining $N$ columns. Subsequently, using that $M_1$ is non-singular, the user computes the error vector $e_i$ as
    \begin{align*}
        e_i = V_i'(D)-V_i'(U)M_1^{-1}M_2.
    \end{align*}
    It is noteworthy that the error vector $e_i$ depends solely on the noise matrices and the database files. Furthermore, the information extraction process only requires knowledge of the matrices $M_1$, $M_2$, and $\Delta$, thereby obviating the need for the user to store the matrices $P_1,\ldots,P_n$ at high cost.
    \item [3.] For each entry $e_{i,j}$, $j=1,\ldots,N$ in $e_i$ set 
    \begin{align*}
        e_{i,j}' = \begin{cases}
            p-e_{i,j} & \textrm{ if } e_{i,j}>\frac{p}{2}\\
            e_{i,j} & \textrm{ else. }
        \end{cases}
    \end{align*}
    \item [4.] Calculate 
    \begin{align*}
        e_{i,j}'' = \begin{cases}
            e_{i,j}'-(e_{i,j}\hspace{-0.2cm}\mod q) & \textrm{ if } (e_{i,j}'\hspace{-0.2cm}\mod q) < \frac{q}{2}\\
            e_{i,j}'- (e_{i,j}\hspace{-0.2cm}\mod q) + q &\textrm{ else}
        \end{cases}
    \end{align*}
    for every $j=1,\ldots,N$.
    \item [5.] Given that the elements of the file $A_{i_0}$ are related to the error vector $e_{i,j}''$ by $A_{i_0}[i,j]=\frac{e_{i,j}''}{q}$ for every $j=1,\ldots,N$, the user can reconstruct the file $A_{i_0}$ by iterating over all rows $V_i$ of $R$.\\
\end{itemize}

Although the authors don't claim particular security levels, they analyze the scheme's vulnerability to lattice and structural attacks. For practical implementation, they finally recommend using the parameters $l_0=20$, $q=2^{39}$, $N=50$, $p=2^{60}+325$, and a maximum database size of $n \leq 10,000$.  As intended, the authors demonstrate that the PIR scheme can be viewed as a trade-off between communication costs and computational costs.\\

\section{The first attack}

As demonstrated by the authors in \cite{Attack}, a linear relationship between the query $B = (B_1, \ldots, B_n)^T$ and the noise matrices $\epsilon = (\epsilon_1, \ldots, \epsilon_n)^T$ can be established. This linear relationship reveals a vulnerability in the scheme, particularly for small databases. Since our attack builds upon their approach, we provide a brief summary of their method. \\
Consider the lattice
\begin{align*}
    L_p(B) =\{y\in\mathbb{Z}^{nN} : y = Bx\hspace{-0.2cm} \mod p \textrm{ for some }x\in\mathbb{Z}_p^{2N}\}. 
\end{align*}
The structural properties of $B$ enabled the authors to demonstrate \cite[Theorem 1]{Attack}, that the columns of $\epsilon$ are themselves contained within the lattice. This, in turn, implies the existence of a bridging matrix $D \in M_{2N \times N}(\mathbb{Z}_p)$, such that
\begin{align}\label{bridging}
   \begin{pmatrix}
        B_1\\
        \vdots\\
        B_n
    \end{pmatrix} D = \begin{pmatrix}
        \epsilon_1\\
        \vdots\\
        \epsilon_n
    \end{pmatrix}
\end{align}
with the multiplication performed in $\mathbb{Z}_p$. The entries of the columns of $\epsilon$ are restricted to values in $\{-1,1\}$. The only exception is the diagonal entries of $\epsilon_{i_0}$, which take values in $\{-q,q\}$. Given that $q$ is significantly smaller than $p$, the columns of $\epsilon$ can be characterized as short lattice vectors. \\

Initially, it may seem intuitive to apply lattice reduction techniques directly to the lattice $L_p(B)$. However, a closer examination reveals that this approach is impractical due to the lattice's high rank, which renders the computational requirements prohibitively expensive.
Instead of directly manipulating the original lattice, the authors of \cite{Attack} introduce a sequence of reduced-dimension lattices, denoted by $L_i$, to make the problem computationally manageable. Thereby, the lattices $L_i$ are defined by 
\begin{align*}
L_i = \{y\in\mathbb{Z}^{2N+k} : y = C_ix \hspace{-0.2cm}\mod p  \textrm{ for some }x\in\mathbb{Z}_p^{2N}\},
\end{align*}
where the matrices $C_i\in\mathbb{Z}_p^{(2N+k)\times 2N}$ are consecutive rows of $B$ formed as 
\begin{align*}
    C_i =\begin{pmatrix}
        B_i \\
        B_{i+1}\\
        B_{i+2}|_{[1:k,:]}
    \end{pmatrix}
\end{align*}
for every $i=1,\ldots,n$ and an attack parameter $k\in\{1,\ldots,N\}$. To ensure the well-definedness of the equation, set $B_{i+n} = B_i$.

 From (\ref{bridging}), we directly obtain
\begin{align}\label{NewEpsilon}
    C_i D = \begin{pmatrix}
        \epsilon_i\\
        \epsilon_{i+1}\\
        \epsilon_{i+2}|_{[1:k,:]}
    \end{pmatrix}  = \overline{\epsilon_i}
\end{align}
for every $i=1\ldots,n$. To facilitate the recovery of the index of interest, the authors formulated a CVP on the reduced-dimension lattices $L_i$. Specifically, they investigated CVP instances with target vectors $t_j = (0, \ldots, 0, q, 0, \ldots, 0) \in \mathbb{Z}^{2N+k}$ for $j = 1, \ldots, N$, where the value $q$ is located in the $j$-th position. The value $q$ in the target vector aligns with the amplified diagonal entries of the noise matrix $\epsilon_{i_0}$, enabling the adversary to distinguish the desired index $i_0$.  \\ 

Based on the provided definition, the authors demonstrated \cite[Theorem 2]{Attack} that, when considering the lattice $L_{i_0}$, the closest vector to $t_j$ is, apart from the sign, the $j$-th column of $\overline{\epsilon_{i_0}}$ as given in (\ref{NewEpsilon}) with high probability. However, the columns of $\overline{\epsilon_{i_0}}$ have a specific structure: the elements are restricted to $\{-1,1\}$, except for values corresponding to the diagonal entries of $\epsilon_{i_0}$, which are $\pm q$. Due to this highly non-random structure, we can distinguish $L_{i_0}$ from other lattices $L_i$, $i \neq i_0$, with high probability by analyzing the closest lattice vector. \\

To solve the CVP, the authors utilized Kannan's embedding technique \cite{Kannan}. Initially, a short lattice basis $S$ of $L_i$ was constructed. Using this short basis, an embedded lattice
\begin{align*}
L_{i,j} =\left\{ y\in\mathbb{Z}^{2N+k+1} : y= \begin{pmatrix}
S & \; t_j \\
0 &\; 1
\end{pmatrix}x \textrm{ for some }x\in\mathbb{Z}^{2N+1} \right\}
\end{align*}
is defined. As discussed in Section \ref{Preliminaries}, solving the SVP on $L_{i,j}$ yields the solution to the CVP on $L_i$ with respect to $t_j$ if the parameters are appropriately chosen. \\

To summarize, the attack proceeds as follows. Initially, we iterate over all lattices $L_{i}$ and, within each lattice, over all target vectors $t_j$. For each target vector, we solve the CVP as an SVP on the embedded lattice $L_{i,j}$. If the solution exhibits the specified structural properties, we identify the current index $i$ as the desired file index, thereby recovering the target file. Within this attack, the choice of $k$ represents a trade-off between computational cost, which increases with the value of $k$, and the growing gap between the shortest and second shortest vector within the embedded lattice, which also increases with $k$. This larger gap facilitates the more efficient solution of the SVP.

\begin{Remark}\label{TargetVector}
    Instead of restricting the non-zero entry in $t_j$ to the first $N$ out of the $2N+k$ positions, an alternative strategy could involve varying the position of $q$ in $t_j$ across all possible positions. This approach would entail examining a larger number of target vectors. Meanwhile, we only need to study a fraction of all lattices $L_i$. While the overall number of SVPs to be solved is similar for both approaches, the complexity of this approach is slightly better. The reason behind this improvement is the ability to reuse the short lattice basis $S$ of $L_i$ for each target vector, thereby enhancing the computational efficiency of the algorithm. 
\end{Remark}

\begin{Remark}\label{tj}
    It is important to note that, from a mathematical perspective, iterating over all target vectors $t_j$, $j=1,\ldots,N$ is not required since solving the CVP exactly with the first target vector would already yield the desired result. Nevertheless, lattice reduction algorithms are inherently approximative in nature. By iterating through multiple target vectors $t_j$, the risk of obtaining a false result is mitigated, thereby enhancing the reliability of the algorithm.
\end{Remark}

In \cite{Attack}, the authors empirically validated the efficacy of their proposed attack through numerical experiments conducted on a database with $n=9$ files. However, there is a linear dependency between the number of files in the database and the number of lattice reduction needed for the attack. More specifically, even if only one target vector $t_j$ is considered for each lattice $L_i$ as suggested in Remark \ref{tj}, the average number of CVPs that need to be solved is $\frac{n}{2}$, with the worst-case scenario requiring $n$ CVPs. Taking into account the large computational time for lattice reduction algorithms, this linear dependency makes the attack impractical for databases exceeding a certain size.

\section{Improved attack}\label{ImprovedAttack}
While the attack strategy outlined in the previous section is valid for small databases, its scalability is compromised by a time complexity that grows linearly with the number of files in the database. To address this issue, we introduce a two-stage approach that combines the original attack with a preprocessing step designed to decrease the number of blocks in the matrix $B = (B_1, \ldots, B_n)$. By iteratively eliminating blocks $B_i$ that do not match the block $B_{i_0}$, we generate a condensed query matrix that can be efficiently processed using the original attack, thus extending its applicability to larger databases.\\

To obtain a logarithmic instead of a linear complexity in $n$, we proceed as in a binary search and initially split $B = (B_1, \ldots, B_n)$ into two parts
\begin{align*}
    B|_{[1:3lN,:]} =\begin{pmatrix}
        B_{1}\\
        \vdots\\
        B_{3l}
    \end{pmatrix} \textrm{ and } B|_{[3lN+1:nN,:]}=\begin{pmatrix}
        B_{3l+1} \\
        \vdots\\
        B_{n}
    \end{pmatrix} ,
\end{align*}
where $l=\lceil \frac{n}{6}\rceil$. By selecting $l$ in this manner, we ensure that each part contains approximately $\frac{n}{2}$ of the $n$ matrices $B_i$. In the subsequent analysis, we first examine the matrix $B|_{[1:3lN,:]}$ to determine whether $B_{i_0}$ is included within this part of $B$. If $i_0$ is found to be within the range $\{1,\ldots,3l\}$, we proceed with the first matrix and disregard the second; otherwise, we move to the second matrix and disregard $ B|_{[1:3lN,:]}$.\\

To determine whether $i_0$ is included in the first $3l$ indices, we further divide the matrix $ B|_{[1:3lN,:]}$ into blocks of three matrices each and sum those blocks up such that we obtain
\begin{align}\label{H}
    H = \begin{pmatrix}
        H_1 \\
        H_2 \\
        H_3 
    \end{pmatrix} = \sum_{i=0}^{l-1} \begin{pmatrix}
        B_{1+3i}\\
        B_{2+3i}\\
        B_{3+3i}
    \end{pmatrix}. 
\end{align}
Summing blocks reduces the number of candidate matrices while preserving the overall structure, which can be illustrated as follows. Adding two matrices $B_i$ and $B_j$ results in
\begin{align*}
    B_i+B_j &= [(P_i+P_j)M_1\, | \, (P_i+P_j)M_2 +(\epsilon_i+\epsilon_j)]\Delta
\end{align*}
for $i,j\in\{1,\ldots,n\}$. Given that $P_i+P_j$ is again a random matrix, the structure of the sum remains similar to the original, with the exception of the noise term. The noise matrix $\epsilon_i+\epsilon_j$ in the sum is now characterized by entries in $\{-2,0,2\}$, where each of $\pm 2$ occurs with a probability of $\frac{1}{4}$ and $0$ occurs with a probability of $\frac{1}{2}$. Generalizing this to the summation of $l$ matrices yields  
\begin{align*}
    H_j = [\tilde{P}_j M_1\, | \, \tilde{P}_j M_2 + \gamma_j]\Delta
\end{align*}
for $j=1,2,3$ with 
\begin{align*}
    \tilde{P}_j = \sum_{i=0}^{l-1} P_{j+3i}
\end{align*}
being a random matrix. The noise matrix $\gamma_j$ is given by
\begin{align*}
    \gamma_j = \sum_{i=0}^{l-1} \epsilon_{j+3i}
\end{align*}
for $j=1,2,3$, such that the entries of $\gamma_j$ follow a shifted binomial distribution either on $\{-l,\ldots,l\}\cap (2\mathbb{Z})$ if $l$ is even or on $\{-l,\ldots,l\}\cap (2\mathbb{Z}+1)$ if $l$ is odd.
We can directly transfer the bridging relation (\ref{bridging}) to the summed matrices as 
\begin{align}\label{Bridge}
    HD = \begin{pmatrix}
        \gamma_1\\
        \gamma_2\\
        \gamma_3
    \end{pmatrix}=\gamma.
\end{align}
Study the lattice defined by the summed matrix $H$ as 
\begin{align}\label{Lattice}
    L_p(H) = \{ y \in\mathbb{Z}^{3N}\hspace{-0.15cm}: y = Hx\hspace{-0.2cm}\mod p  \textrm{ for some }x\in\mathbb{Z}_p^{2N}\}. 
\end{align}
From the bridging relation (\ref{Bridge}), we directly obtain the following relation between $\gamma$ and the lattice $L_p(H)$. 
\begin{Lemma}
    Let $L_p(H)$ denote the lattice defined by (\ref{Lattice}), where $H$ is derived from the summation of submatrices of $B$ as specified in (\ref{H}). Then each column of $\gamma = (\gamma_1, \gamma_2, \gamma_3)^T$ represents a lattice vector within $L_p(H)$.
\end{Lemma}

Note that the columns of $\gamma$ are not only lattice vectors but also short lattice vectors when $l$ is small, since the entries of $\gamma$ are confined to a subset of $\{-l, \ldots, l\}$. In analogy to the original attack, our objective is to solve a CVP on the lattice $L_p(H)$. To this end, we define a set of target vectors $t_j = (0, \ldots, 0, q, 0, \ldots, 0)$ for $j = 1, \ldots, 3N$, where the value $q$ is placed at the $j$-th position. It is worth noting that, as highlighted in Remark \ref{TargetVector}, our approach employs a larger set of target vectors, yet reduces the number of lattices utilized compared to the original attack. This modification is motivated by an improvement in computational efficiency. \\

The subsequent theorem establishes that, for sufficiently small values of $l$, it is possible to decide whether $B_{i_0}$ is contained in $H$ or not.
\begin{Theorem}\label{NewBound}
    Suppose that the number of additions $l$ is bounded from above such that 
    \begin{align*}
        N_{\sqrt{3lN}}(0)\leq p^{N-2},
    \end{align*}
    where $N_R(x)$ is the number of integer points within the $3N$-dimensional sphere as defined in Section \ref{Preliminaries}. Assume that the desired file index $i_0$ is an integer such that $i_0 \in \{1, \ldots, 3l\}$. Then there is an $a\in\{1,2,3\}$ such that $B_{i_0}$ is a summand in $H_a$. Let $L_p(H)$ be the lattices as defined in (\ref{Lattice}) and consider the target vector $t_j$, where the index $j$ satisfies
  \begin{align*}
        (a-1)N <  j \leq  aN.
    \end{align*} 
In other words, we are considering a target vector with a non-zero entry within the same block as $B_{i_0}$. Then 
\begin{align*}
u=j-(a-1)N\in\{1,\ldots,N\}    
\end{align*}
is the row within block $a$, in which $t_j$ has the non-zero entry. If the corresponding diagonal entry in $\epsilon_{i_0}$ equals $\epsilon_{i_0}[u,u]=q$, then with probability at least 
\begin{align*}
    \exp\left(-p^{-2}\frac{p^{3N}}{p^{3N}-1}\right),
\end{align*}
the closest lattice vector with respect to $t_j$ is given by $\gamma_{[:,u]}$. Conversely, if $\epsilon_{i_0}[u,u]=-q$ then with the same probability, the closest lattice vector with respect to $t_j$ is given by $-\gamma_{[:,u]}$.
\end{Theorem}

\begin{Remark}
   The formulation of Theorem \ref{NewBound} is quite technical, such that we want to give a more accessible interpretation before proceeding with its proof. From an intuitive perspective, it is evident that the number of additions $l$ must be bounded from above. If $l$ would be unbounded, the difference between $\epsilon_{i_0}$ with $\pm q$ on its diagonal and all other $\epsilon_i$ becomes negligible in the summed matrix $H$. The bound given in the theorem is implicit, requiring that $l$ is chosen such that the number of integer points inside a $3N$-dimensional sphere of radius $\sqrt{3Nl}$ centered at the origin is at most $p^{N-2}$. Utilizing the approximation on $N_R(x)$ from Section \ref{Preliminaries}, this implicit definition can be translated into an approximate bound on $l$. In detail, we obtain the upper bound $l_{max}$ of $l$ as
   \begin{align*}
       l_{max} &\approx \frac{\Gamma\left(\frac{3N}{2}+1\right)^{\frac{2}{3N}}}{3N\pi} p^{\frac{2}{3}(1-\frac{2}{N})}\\
   & \approx \frac{1}{2\pi e} p^{\frac{2}{3}(1-\frac{2}{N})},  
   \end{align*}
   using that 
   \begin{align*}
       \lim_{x\to\infty} \frac{\sqrt[x]{\Gamma(x+1)}}{x} \to e^{-1}.
   \end{align*}
  The initial paper proposed values of $p=2^{60}+325$ and $N=50$, yielding an estimated maximum length $l_{max} \approx 2.12 \cdot 10^{10}$. Notably, our approach involves adding at most $\lceil \frac{n}{6}\rceil$ blocks. Given the constraint on database size, where $n\leq 10,000$, it is evident that in this case $l$ significantly falls below $l_{max}$. Consequently, for practical applications and parameters as suggested in \cite{OriginalScheme}, the assumption on $l$ can be readily fulfilled. \\
 Consider a matrix $H$ that incorporates $B_{i_0}$. According to Theorem \ref{NewBound}, for a suitably chosen target vector $t_j$, the closest vector to $t_j$ is given by a particular column of $\gamma$ with probability
    \begin{align*}
    \exp\left(-p^{-2}\frac{p^{3N}}{p^{3N}-1}\right)\approx 1-p^{-2}.
    \end{align*}
    Substituting the proposed value $p=2^{60}+325$ into this expression yields an approximate probability of $(1-7.52\cdot 10^{-37})$ that the closest vector is indeed given by this column of $\gamma$. Although the probability obtained is already remarkably high, it is reasonable to expect an even higher probability of success in practical applications. This is because the proof is based on a worst-case analysis, considering the maximum possible value of $l$ and the maximum possible length of the column of $\gamma$. In reality, the actual values of these parameters are likely to be more favorable, leading to an even higher probability of success. Consequently, it is highly unlikely that this theorem will ever fail in practical applications.
    
\end{Remark}

\begin{proof}
Initially, let us examine the vector $\gamma_{[:,u]}$. As a result of the summation, for every $v \in \{1, \ldots, 3N\} \setminus \{j\}$, the $v$-th entry of this vector satisfies 
\begin{align*}
    \gamma_{[v,u]} \in \{-l,\ldots,l\}.
\end{align*}
In contrast, when $v = j$ the corresponding entry is constrained to the range
\begin{align*}
     \gamma_{[j,u]} \in \{-l+1,\ldots,l-1\} \pm q
\end{align*}
such that
\begin{align*}
    ||\pm \gamma_{[:,u]}-t_j|| &\leq \sqrt{(3N-1)l+(l-1)}\\
    &\leq \sqrt{3Nl}. 
\end{align*}

Note that, due to the entries of $\gamma$ being distributed according to a shifted binomial distribution, the length of the columns will be even shorter with high probability. \\

As a subsequent step, we demonstrate that, with high probability, there exists no lattice vector with a distance to the target vector that is equal to or shorter than this threshold. To this end, let $R>0$ be fixed and recall that $B_R(x)$ denotes the $3N$-dimensional sphere of radius $R$ centered at $x\in\mathbb{R}^{3N}$. We establish a lower bound for the probability that no lattice vector has a distance to $t_j$ that is less than or equal to $R$, which can be expressed as
\begin{align*}
    &P(\nexists\, y\in L_p(H) : ||y-t_j|| \leq R)\\
    =&P(\nexists\, y\in B_R(t_j)\cap L_p(H))\\
    =& P(\nexists\, y\in B_R(t_j)\cap \mathbb{Z}^{3N}:  y= Hx\hspace{-0.2cm}\mod p \textrm{ for some }x\in\mathbb{Z}_p^{2N}),
\end{align*}
where we used the definition of the lattice $L_p(H)$ in (\ref{Lattice}). Due to the randomness of $H$, we obtain
\begin{align*}
    P(\nexists\, y\in L_p(H) : ||y-t_j|| \leq R)
    = & \prod_{y\in B_R(t_j)\cap\mathbb{Z}^{3N}} P(\nexists\, x\in\mathbb{Z}_p^{2N}: y= Hx\hspace{-0.2cm}\mod p ) \\
    =&  \prod_{y\in B_R(t_j)\cap\mathbb{Z}^{3N}} \prod_{x\in\mathbb{Z}_p^{2N}} (1-P(y=Hx \hspace{-0.2cm}\mod p) ).
\end{align*}
For a random choice of the matrix $H\in\mathbb{Z}_p^{3N\times 2N}$ and a random vector $x\in\mathbb{Z}_p^{2N}$, the product $Hx$ is randomly distributed in $\mathbb{Z}_p^{3N}$ such that 
\begin{align*}
    P(y=Hx \hspace{-0.2cm}\mod p) =p^{-3N}. 
\end{align*}
It follows that the probability of no lattice vector being within a distance of $R$ or less from $t_j$ is equal to
\begin{align*}
    P(\nexists\,  y\in L_p(H) : ||y-t_j|| \leq R)
    = &  (1-p^{-3N})^{p^{2N}N_R(t_j)} \\
    = & \exp\left(p^{2N}N_R(t_j) \log (1-p^{-3N}) \right).
\end{align*}
Due to our assumption, for the radius $R=\sqrt{3Nl}$ we obtain 
  \begin{align*}
        N_{\sqrt{3lN}}(t_j)\leq p^{N-2}.
    \end{align*}

For the overall probability that no lattice vector with distance shorter or equal to $R$ exists, this yields
\begin{align*}
    P(\nexists\,  y\in L_p(H) : ||y-t_j|| \leq \sqrt{3Nl})
    \geq & \exp\left(p^{-2+3N}\log(1-p^{-3N})\right)\\
    = & \exp\left(-p^{-2+3N}\sum_{k=1}^{\infty} \frac{p^{-3Nk}}{k}\right)
\end{align*}
using the series representation of $\log(1-x)$. Finally, 
\begin{align*}
    P(\nexists\,  y\in L_p(H) : ||y-t_j|| \leq \sqrt{3Nl})
    \geq  & \exp\left(-p^{-2+3N}\sum_{k=1}^{\infty} p^{-3Nk}\right)\\
    =& \exp\left(-p^{-2}\frac{p^{3N}}{p^{3N}-1}\right), 
\end{align*}
which finishes the proof. \qed
\end{proof}

Under the assumptions of Theorem \ref{NewBound}, the closest lattice vector with respect to $t_j$ exhibits a distinctive pattern with high probability. Specifically, dependent on $l$, the entries of this vector are either all even or all odd, with the notable exception of the $j$-th entry, which possesses the opposite parity. Remarkably, the occurrence of such a configuration is exceedingly rare in random instances, such that this behavior of the closest lattice vector can serve as the foundation for our attack.\\ 

 As in \cite{Attack}, we use Kannan's embedding technique to solve the CVP on $L_p(H)$. Therefore, let $S\in\mathbb{Z}^{3N\times 3N}$ be a short basis of $L_p(H)$ and let the embedded lattice be given by
\begin{align*}
    L_{H,j} =\left\{ y \in\mathbb{Z}^{3N+1} : y=\begin{pmatrix}
        S &\; t_j\\
        0 &\;  M 
    \end{pmatrix}x \textrm{ for some } x\in\mathbb{Z}^{3N+1}\right\}.
\end{align*}

By Theorem \ref{NewBound}, if $B_{i_0}$ is contained in $H$ and the index of the target vector is appropriately selected, the shortest vector is effectively determined by a column of $\gamma$. Building upon this result, our proposed attack proceeds as follows. Initially, we partition the matrix $B$ into two segments and compute the sum of the first segment to obtain the matrix $H$. We then define the lattice $L_p(H)$ and calculate a short basis $S$ of $L_p(H)$. To mitigate computational complexity, as argued in Remark \ref{tj}, we restrict our attack to the target vectors $t_1$, $t_{N+1}$, and $t_{2N+1}$. For these three target vectors, we solve the SVP on the embedded lattice. If the solution exhibits a structural consistency with a column of $\gamma$, we can infer that $B_{i_0}$ is contained in $H$. Furthermore, by retaining the target vector that yields the successful outcome, we can even identify the specific block of $H$ in which $B_{i_0}$ is contained. This enables us to not only disregard the second half of the matrix $B$ but also eliminate the two blocks of $H$ that do not contain $B_{i_0}$, thereby reducing the number of blocks to be considered in the subsequent iteration to $l$. Conversely, if $B_{i_0}$ is not present in $H$, we can deduce that it is contained in the second half of $B$, and the procedure continues with this half. In average, this reduces the number of matrices to be considered by a factor of $\frac{1}{3}$. By recursively applying this approach, we can further narrow down the number of matrices until it reaches a manageable size for the original attack. The pseudocode presented in Algorithm \ref{Algorithm} provides a concise summary of the proposed attack.\\

\begin{figure}[htbp]
\centering
\begin{tikzpicture}
     \begin{axis}[
         width=9cm, 
         height=8cm, 
         xlabel={Database size $n$},
         ylabel={Number of CVPs solved},
         ylabel style={yshift=-8pt},
         xmin=0, xmax=10000, 
         ymin=0, ymax=40, 
         xtick={0,2000,4000,6000,8000,10000}, 
         ytick={0,10,20,30,40}, 
         scaled x ticks = false,
         legend pos=south east,
         grid=both, 
         grid style={dashed, gray!30}, 
         tick align=outside, 
         tick style={black}, 
         legend cell align={left}, 
         legend style={font=\small}, 
         label style={font=\small}, 
         title style={font=\bfseries\footnotesize}, 
     ]
         \addplot[
             thick,
             color=blue,
             mark=*,
             mark options={scale=0.8},
         ] table {
             x y
             100 16
             500 23
             1000 25
             1500 27
             2000 29
             2500 28
             3000 30
             3500 30
             4000 31
             4500 33
             5000 32
             5500 31
             6000 33
             6500 35
             7000 34
             7500 36
             8000 35
             8500 34
             9000 36
             9500 34
             10050 35
         };
         \addlegendentry{Worst Case}
         \addplot[blue, dotted,thick,domain=100:10000, 
    samples=1000, ] {3*(ln(x)-ln(6))/ln(2)+3};
         \addlegendentry{Worst Case Approximation}
          \addplot[
             thick,
             color=red,
             mark=square*,
             mark options={scale=0.8},
         ] table {
             x y
             100 9.5
             500 13.004
             1000 14.512
             1500 15.402
             2000 15.962
             2500 16.66
             3000 16.703
             3500 17.138
             4000 17.3065
             4500 17.764
             5000 18.1946
             5500 17.8349
             6000 18.1075
             6500 18.2863
             7000 18.5796
             7500 18.602
             8000 18.5998
             8500 18.898
             9000 19.083
             9500 19.3997
             10050  
         };
         \addlegendentry{Average Case}
        \addplot[red, dotted,thick,domain=100:10000, 
        samples=1000, ] {2.5*(ln(x)-ln(6))/ln(3)+3};
         \addlegendentry{Average Case Approximation}
     \end{axis}
\end{tikzpicture}
\caption{Exact and approximate number of CVPs to be solved in the worst case and the average case for different database sizes $n$ and a threshold value of $t=6$.} \label{NumberCVPs} 
\end{figure}
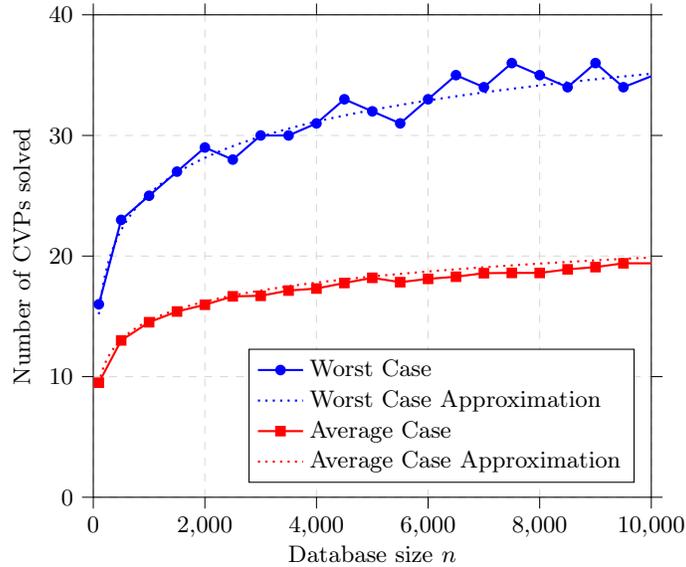

The original attack becomes impractical for large databases because the number of CVPs to solve scales linearly with $n$. Specifically, in the worst-case scenario, the classical attack requires solving $n$ CVPs, while on average, it necessitates solving $\frac{n}{2}$ CVPs. Our improved approach exhibits a logarithmic dependence on $n$, making the attack suitable for all database sizes. \\
To underscore the low complexity of our improved attack, we examine both the worst-case and average-case scenario. Notably, the worst-case scenario corresponds to the instance where $i_0=n$, necessitating the solution of three CVPs in each iteration without yielding a successful outcome, thereby proceeding with the latter half of the matrix. Let $t$ denote the threshold value that triggers the original attack. A theoretical approximation of the number of CVPs to be solved is given by
\begin{align}\label{WC}
    \# \text{CVPs}_{\text{WorstCase}} \approx 3 \frac{\log(n)-\log(t)}{\log(2)} + 0.5t. 
\end{align}
In contrast, for the average case, we derive
\begin{align}\label{AC}
    \#\text{CVPs}_{\text{AverageCase}} \approx 2.5 \frac{\log(n)-\log(t)}{\log(3)} + 0.5t. 
\end{align}
To illustrate the practical implications for the complexity, we consider a threshold value of $t=6$ and plotted the exact as well as the approximate number of CVPs to be solved in our improved attack for various database sizes $n$ in Figure \ref{NumberCVPs}. The variability within the exact worst-case curve can be attributed to the interplay between precomputing and original attack. As the size of $n$ fluctuates, the original attack is initiated with matrices of varying dimensions, resulting in a non-monotonic curve. In contrast, the averaging process in the average case largely mitigates the impact of disparate starting lengths for the original attack, thereby smoothing the curve. The approximate formulas presented in (\ref{WC}) and (\ref{AC}) demonstrate a satisfactory level of precision in approximating the actual number of CVPs for both scenarios.

\begin{algorithm}\label{Algorithm}
 \caption{Improved attack on the lattice-based PIR scheme}
 \KwData{Matrix $B=(B_1,\ldots,B_n)$}
 \KwResult{Index of interest $i_0$}
 Preprocessing: \\
 \While{ $|RemainingBlocks |\geq \, Threshold$}{
  split $RemainingBlocks$ into two parts;\\
    \hspace{0.5cm} $Part1$ = First $3l$ blocks\\
    \hspace{0.5cm} $Part2$ = Remaining blocks\\
  \For{$Part1$}{
        sum up $Part1$ to obtain $H$ \\
        define lattice $L_p(H)$\\
        find a short basis $S$ for $L_p(H)$\\
        \For{each $t_j$}{
            define the embedded lattice $L_{H,j}$\\
            solve the SVP on $L_{H,j}$ \\
            \If{closest vector is valid}{
                $found$=TRUE \\
                break
            }
        }  
    }
    \uIf{$found$=TRUE}{
    update $RemainingBlocks$ to include only the blocks of $Part1$ related to $t_j$
  }
  \Else{
    $RemainingBlocks=Part2$
  }
}
 start the original attack with $RemainingBlocks$
\end{algorithm}

\section{Numerical results}
Using SageMath \cite{sage}, we implemented the attack to display its efficiency. Following the parameters suggested in \cite{OriginalScheme}, we set $l_0=20$, $q=2^{2l_0-1}$, $N=50$, and $p = 2^{60} + 325$. The suggested database size for these parameters is $n \leq 10,000$. We continued the preprocessing until $t=6$ or fewer blocks were left. Subsequently, we initiated the original attack using $k = 34$ as recommended in \cite{Attack}. For all lattice reductions, we employed the LLL algorithm. For the embedded lattice, we set $M = 1$ as the embedding factor. This choice ensures that the gap between the shortest and the second shortest lattice vector is large. Meanwhile, it also increases the probability of false results. However, for this particular case, the procedure operates correctly, leading us to retain this parameter choice. The results are presented in Table \ref{NumericalResults}. We display the minimal and maximal run time of the attack as well as the success rate, which is measured across $100$ trials. 
All calculations were conducted on a laptop equipped with an Intel Core i7-1370P processor with 16 GB RAM. Regardless of the database size, the index of interest can be recovered reliably within minutes, demonstrating the complete breakdown of the scheme. For an easier comparison, the computational process is not executed in parallel. However, the described approach is particularly well-suited for parallel computing. By dividing the query matrix into separate blocks and searching within these blocks in parallel, the computational time can be further reduced.

\begin{table}[]
    \centering
     \caption{Minimal and maximal attack time as well as success rate over $100$ trials for our improved attack with different database sizes $n$ using $l_0=20$, $q=2^{2l_0-1}$, $N=50$ and $p = 2^{60} + 325$. The original attack is triggered at $t=6$ with $k=34$.}
    \begin{tabular}{@{}lccc@{}}
       \toprule
      \multirow{2}{*}{n}   & \; min. attack \;& \;max. attack \; & \;success\;  \\
     &   time (min)&time (min)& \; rate (\%) \;\\
    \midrule
    \midrule
     100   & 2.1 & 7.4 & 100 \\
     1,000  & 3.6  & 12.3 & 100 \\
     5,000  & 4.2 & 17.8 & 100 \\
     10.000 & 5.4 & 18.6 & 100 \\
     \bottomrule
    \end{tabular}
    \label{NumericalResults}
\end{table}

\section{Conclusion}
In this paper, we identified a critical vulnerability in the lattice-based Private Information Retrieval (PIR) scheme proposed by Melchor and Gaborit. Our contribution builds upon the previous work by Liu and Bi, which was restricted to databases of specific sizes, and extends it to databases of arbitrary sizes. By employing a binary-search-like approach to presort the matrices, we substantially reduce the number of lattice problems that need to be considered. This optimization reduces the complexity of the attack from linear to logarithmic in $n$, rendering the scheme insecure for all parameter choices. Additionally, our numerical evaluations underline the feasibility and reliability of our approach, with successful attacks executed within minutes on standard hardware.

\subsubsection{\ackname} 
This work has been supported by funding from Agentur für Innovation in der Cybersicherheit GmbH.

%
%
%
 \bibliographystyle{splncs04}
%


\begin{thebibliography}{10}
\providecommand{\url}[1]{\texttt{#1}}
\providecommand{\urlprefix}{URL }
\providecommand{\doi}[1]{https://doi.org/#1}

\bibitem{Ajtai}
Ajtai, M.: The shortest vector problem in {$L_2$} is {NP}-hard for randomized
  reductions (extended abstract). In: Proceedings of the Thirtieth Annual ACM
  Symposium on Theory of Computing (STOC). pp. 10 -- 19. ACM Press, Dallas,
  Texas, United States (1998). \doi{10.1145/276698.276705}

\bibitem{Chamizo1995}
Chamizo, F., Iwaniec, H.: On the sphere problem. Revista Matemática
  Iberoamericana  \textbf{11}(2),  417 -- 429 (1995). \doi{10.4171/RMI/178}

\bibitem{PIR}
Chor, B., Kushilevitz, E., Goldreich, O., Sudan, M.: Private information
  retrieval. Journal of the ACM (JACM)  \textbf{45}(6),  965 – 981 (1998).
  \doi{10.1145/293347.293350}

\bibitem{NPcomplete}
van Emde-Boas, P.: Another NP-complete partition problem and the complexity of
  computing short vectors in a lattice. Report, Department of Mathematics,
  University of Amsterdam (1981)

\bibitem{Kannan}
Kannan, R.: Minkowski's convex body theorem and integer programming.
  Mathematics of Operations Research  \textbf{12}(3),  415 -- 440 (1987).
  \doi{10.1287/moor.12.3.415}

\bibitem{Attack}
Liu, J., Bi, J.: Cryptanalysis of a fast private information retrieval
  protocol. In: Proceedings of the 3rd ACM International Workshop on Asia
  Public-Key Cryptography (AsiaPKC '16). pp. 56 -- 60. Association for
  Computing Machinery, New York, NY, USA (2016). \doi{10.1145/2898420.2898427}

\bibitem{Odlyzko1990}
Mazo, J., Odlyzko, A.: Lattice points in high-dimensional spheres. Monatshefte
  für Mathematik  \textbf{110}(1),  47 -- 62 (1990). \doi{10.1007/BF01571276}

\bibitem{OriginalScheme}
Melchor, C., Gaborit, P.: A fast private information retrieval protocol. In:
  Proceedings of the IEEE International Symposium on Information Theory (ISIT),
  Toronto, ON, Canada. pp. 1848 -- 1852 (2008). \doi{10.1109/ISIT.2008.4595308}

\bibitem{sage}
Stein, W., et~al.: {S}age {M}athematics {S}oftware ({V}ersion 10.3). The Sage
  Development Team (2024), {\tt http://www.sagemath.org}

\bibitem{StudyKannan}
Wang, Y., Aono, Y., Takagi, T.: An experimental study of {K}annan's embedding
  technique for the search {LWE} problem. In: Qing, S., Mitchell, C., Chen, L.,
  Liu, D. (eds.) Information and Communications Security. pp. 541 -- 553.
  Springer International Publishing, Cham (2018).
  \doi{10.1007/978-3-319-89500-0_47}

\end{thebibliography}

\end{document}